\documentstyle[epsfig,12pt]{article}

\begin{document}
\begin{titlepage}
\begin{flushright}
hep-th/0110199 \\ UFIFT-HEP-01-17
\end{flushright}
\vspace{.4cm}
\begin{center}
\textbf{Light Cone QED in a Homogeneous Electric Background$^{\dagger}$}
\end{center}
\begin{center}
R. P. Woodard$^{\ddagger}$
\end{center}
\begin{center}
\textit{Department of Physics \\ University of Florida \\ 
Gainesville, FL 32611 USA}
\end{center}

\begin{center}
ABSTRACT
\end{center}
\hspace*{.5cm} I present an exact solution for the Heisenberg picture, Dirac 
electron in the presence of an electric field which depends arbitrarily upon
the light cone time parameter $x^+ = (t+x)/\sqrt{2}$. This is the largest class
of background fields for which the mode functions have ever been obtained. The
solution applies to electrons of any mass and in any spacetime dimension. The
traditional ambiguity at $p^+ = 0$ is explicitly resolved. It turns out that 
the initial value operators include not only $(I + \gamma^0 \gamma^1) \psi$ at
$x^+ = 0$ but also $(I - \gamma^0 \gamma^1) \psi$ at $x^- = -L$. Pair creation 
is a discrete and instantaneous event on the light cone, so one can compute the
particle production rate in real time. In $D=1+1$ dimensions one can also see
the anomaly. Another novel feature of the solution is that the expectation 
value of the currents operators depends non-analytically upon the background
field. This seems to suggest a new, strong phase of QED.

\begin{flushleft}
\end{flushleft}
\vspace{.4cm}
\begin{flushleft}
$^{\dagger}$ Talk given at the 6th Workshop on Non-Perturbative QCD, Paris, 
France, June 5-9, 2001. \\
$^{\ddagger}$ e-mail: woodard@phys.ufl.edu
\end{flushleft}
\end{titlepage}

\section{Introduction}

I will be reporting on work done with my good friends Nikolaos Tsamis and 
Theodore Tomaras, both from the University of Crete. Our study concerns a 
generalization of Schwinger's classic treatment of QED in the presence of 
an external electromagnetic field \cite{Schwinger}. Recall that Schwinger 
computed what is now known as the one loop effective action for two special 
classes of background fields: the case of constant $F_{\mu\nu}$ and the case 
of a plane wave solution of the free Maxwell equations. What we did is to 
solve the Dirac equation for the electron field operator in the presence 
of an electric background field $\vec{E} = E(x^+) \widehat{x}$ which 
points in the ${\widehat x}$ direction and which depends arbitrarily upon 
the light cone time parameter $x^+ \equiv (t + x)/\sqrt{2}$ \cite{TTW1,TTW2}.
We were originally seeking a system in which back-reaction could be studied
without the complications that attend the analogous problem in the presence
of a background metric. However, the solution turns out to have a number of
interesting applications in its own right which are the subject of this 
report. I will first give the solution and then devote a section to each
application.

We define the light cone coordinates as $x^{\pm} \equiv (t \pm x)/\sqrt{2}$.
The remaining, ($D-2$) transverse coordinates are denoted thusly: $x_{\bot}$. 
We work in the gauge where $A_+ = 0$ and,
\begin{equation}
A_-(x^+) = - \int_0^{x^+} du E(u) \; .
\end{equation}
The transverse components of the vector potential vanish, $A_{\bot} = 0$.
It is useful as well to define $\pm$ spinor components,
\begin{equation}
\psi_{\pm}(x^+,x^-,x_{\bot}) \equiv \frac12 (I \pm \gamma^0 \gamma^1)
\psi(x^+,x^-,x_{\bot}) \; .
\end{equation}
With these conventions the Dirac equation reduces to the following system,
\begin{eqnarray}
i \partial_+ \psi_+ & = & \frac12 \left(m + i \nabla_{\bot} \cdot 
\gamma_{\bot}\right) \gamma^- \psi_- \; , \\
(i \partial_- - e A_-) \psi_- & = & \frac12 \left(m + i \nabla_{\bot} \cdot 
\gamma_{\bot}\right) \gamma^+ \psi_+ \; .
\end{eqnarray}

The Fourier transform on $x^-$ does not properly exist, but it is possible 
to transform on the transverse coordinates as usual,
\begin{equation}
\widetilde{\psi}_{\pm}(x^+,x^-,k_{\bot}) \equiv \int d^{D-2}x_{\bot}
e^{-i k_{\bot} \cdot x_{\bot}} \psi_{\pm}(x^+,x^-,x_{\bot}) \; .
\end{equation}
In this notation the general solution for $\widetilde{\psi}_+$ is 
\cite{TTW2},
\begin{eqnarray}
\lefteqn{\widetilde{\psi}_+(x^+,x^-,k_{\bot}) = \int_{-L}^{\infty}dy^- 
\int_{-\infty}^{+\infty} {{dk^+}\over{2\pi}} e^{i(k^++i/L) (y^- - x^-)} }
\nonumber \\
& & \Biggl\{{\cal E}[eA_-](0,x^+;k^+,k_{\bot}) \widetilde{\psi}_+(0,y^-,k_{
\bot}) - {i\over 2} (m - k_{\bot} \cdot \gamma_{\bot}) \gamma^- \nonumber \\
& & \times \int_0^{x^+} dy^+ e^{-ieA_-(y^+) (y^- + L)} {\cal E}[eA_-](y^+,x^+;
k^+,k_{\bot}) \widetilde{\psi}_-(y^+,-L,k_{\bot}) \Biggr\} \; , \label{psi+}
\end{eqnarray}
where the $E$-dependent mode function is,
\begin{equation}
{\cal E}[eA_-](y^+,x^+;k^+,k_{\bot}) \equiv \exp\Biggl[- {i\over 2} \omega^2_{
\bot} \int_{y^+}^{x^+} {du \over k^+ - eA_-(u) + i/L} \Biggr] \; ,
\end{equation}
and $\omega^2_{\bot} \equiv m^2 + k_{\bot} \cdot k_{\bot}$. The field 
$\widetilde{\psi}_-$ is obtained by differentiating $\widetilde{\psi}_+$,
\begin{equation}
\widetilde{\psi}_-(x^+,x^-,k_{\bot}) = \Biggl({m - k_{\bot} \cdot 
\gamma_{\bot} \over \omega^2_{\bot}} \Biggr) \gamma^+ i \partial_+ 
\widetilde{\psi}_+(x^+,x^-,k_{\bot}) \; . \label{psi-}
\end{equation}

\centerline{\psfig{figure=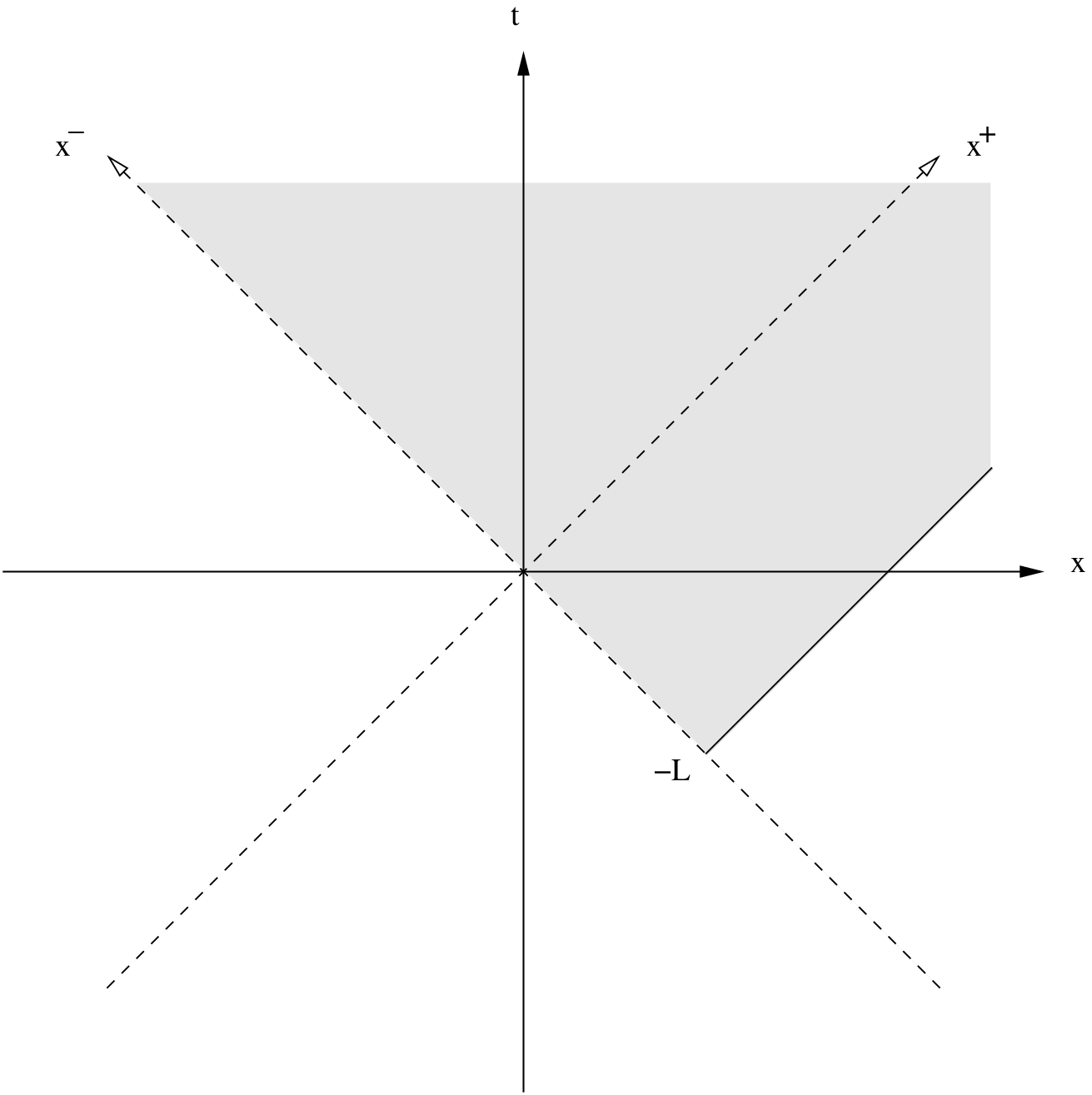,height=3.5in}}

\centerline{Fig.~1: The domain of solution is shaded.}
\vskip 1cm

The domain of our solution is depicted in Fig.~1. It is defined by $x^+ > 0$ 
and $x^- > -L$, for arbitrary $L$. The initial value operators consist of
$\psi_+(0,x^-,x_{\bot})$, for all $x^- > -L$, and $\psi_-(x^+,-L,x_{\bot})$
for all $x^+ > 0$. Canonical quantization reveals the nonvanishing portion
of the anti-commutation algebra to be,\cite{TTW1}
\begin{eqnarray}
\left\{\psi_+(0,x^-,x_{\bot}),\psi_+^{\dagger}(0,y^-,y_{\bot})\right\} & = &
\frac{P_+}{\sqrt{2}} \delta(x^- - y^-) \delta^{D-2}(x_{\bot} - y_{\bot}) 
\; , \quad \\
\left\{\psi_-(x^+,-L,x_{\bot}),\psi_-^{\dagger}(y^+,-L,y_{\bot})\right\} & = &
\frac{P_-}{\sqrt{2}} \delta(x^- - y^-) \delta^{D-2}(x_{\bot} - y_{\bot}) 
\; , \quad
\end{eqnarray}
where $P_{\pm} \equiv (I \pm \gamma^0 \gamma^1)/2$. We assume that the initial
value operators act upon ``the vacuum'' in the same way as they do for zero
electric field. Computing the VEV of any operator therefore consists of first
employing (\ref{psi+}-\ref{psi-}) to express that operator in terms of the 
initial value operators, and then taking the expectation value of these in
the free theory. It is often useful to take the large $L$ limit as well.

\section{Particle production}

By simply deleting the $k^+$ integration from (\ref{psi+}) we obtain a finite
$L$ analog of the Fourier transform on $x^-$,
\begin{eqnarray}
\lefteqn{\widehat{\psi}_+(x^+,k^+,k_{\bot}) \equiv } \nonumber \\
& & \int_{-L}^{\infty}dy^- e^{i(k^++i/L) y^- } \Biggl\{{\cal E}(0,x^+;k^+,
k_{\bot}) \widetilde{\psi}_+(0,y^-,k_{\bot}) - {i\over 2} (m - k_{\bot} \cdot 
\gamma_{\bot}) \gamma^-  \nonumber \\
& & \quad \times \int_0^{x^+} dy^+ e^{-ieA_-(y^+) (y^- + L)} {\cal E}(y^+,x^+;
k^+,k_{\bot}) \widetilde{\psi}_-(y^+,-L,k_{\bot}) \Biggr\} . \label{psihat}
\end{eqnarray}
This operator comes very close to being an eigenfunction of the light cone 
Hamiltonian,
\begin{eqnarray}
\lefteqn{-i {\partial \over \partial x^+} {\widehat \psi}_+(x^+,k^+,k_{\bot}) 
= - {\omega^2_{\bot}/2 \; {\widehat \psi}_+(x^+,k^+,k_{\bot}) \over k^+ - 
e A_-(x^+) + i/L }} \nonumber \\
& & \hspace{1cm} - \frac12 (m - k_{\bot} \cdot \gamma_{\bot}) \gamma^- 
{\widetilde \psi}_-(x^+,-L,k_{\bot}) {i e^{-i (k^+ + i/L) L} \over k^+ - 
e A_-(x^+) + i/L} \; . \label{eigen}
\end{eqnarray}
In the limit of large $L$ the second term contributes only at $k^+ = e
A_-(x^+)$,
\begin{equation}
\lim_{L \rightarrow \infty} 
{i e^{-i (k^+ -e A_-(x^+) + i/L) L} \over k^+ - e A_-(x^+) + i/L} = 2 \pi
\delta\left(k^+ - e A_-(x^+) \right) \; . \label{delta}
\end{equation}
In the limit of infinite $L$ this means that, for $k^+ > e A_-(x^+)$, the 
field (\ref{psihat}) annihilates electrons whereas, for $k^+ < e A_-(x^+)$, it 
must create positrons. To find the amplitude we evaluate its anti-commutator,
\begin{eqnarray}
\lefteqn{\left\{{\widehat \psi}_+(x^+,k^+,k_{\bot}),{\widehat \psi}_+^{
\dagger}(x^+,q^+,q_{\bot})\right\} =} \nonumber \\
& & \hspace{3cm} \frac{P_+}{\sqrt{2}} {i e^{-i (k^+ - q^+ + 2 i/L) L} \over 
k^+ - q^+ + 2 i/L} (2\pi)^{D-2} \delta^{D-2}(k_{\bot} - q_{\bot}) \; . 
\label{amp}
\end{eqnarray}
From (\ref{delta}-\ref{amp}) we see that, in the large $L$ limit, ${\widehat 
\psi}_+(x^+,k^+,k_{\bot})$ creates or destroys particles with amplitude $2^{-
\frac14}$ within the $P_+$ spinor subspace.

Let us assume that the electric field is positive, in which case $eA_-(x^+)$
is a monotonically increasing function of $x^+$. Hence evolution in $x^+$ 
carries more and more electron annihilation operators through the singular
point at which $k^+ = e A_-(x^+)$, after which they become positron creation
operators. The Heisenberg picture vacuum does not change, but since the meaning
of which operator creates a particle changes, so too must our interpretation 
of the vacuum state's occupation number. To find the probability that the
vacuum contains a positron of momentum $k^+$ and $k_{\bot}$ in one of the two
spin states we compute,
\begin{eqnarray}
\lefteqn{\lim_{L \rightarrow \infty} \sqrt{2} \left\langle \Omega \left\vert 
{\widehat \psi}_+^{\dagger}(x^+,k^+,k_{\bot}) {\widehat \psi}_+(x^+,q^+,
q_{\bot}) \right\vert \Omega \right\rangle = } \nonumber \\
& & \left[1 - {\rm Prob}(x^+,k^+,k_{\bot}) \right] \frac{P_+}{\sqrt{2}} 2\pi 
\delta(k^+ - q^+) (2\pi)^{D-2} \delta^{D-2}(k_{\bot} - q_{\bot}) \; .
\end{eqnarray}
The result can be expressed in terms of the functions $X(k^+)$ and 
$\lambda(k^+,k_{\bot})$,
\begin{equation}
k^+ \equiv e A_-\left(X(k^+)\right) \qquad , \qquad \lambda(k^+,k_{\bot}) 
\equiv {\omega_{\bot}^2 \over 2 \vert e E(X(k^+)) \vert} \; .
\end{equation}
The positron creation probability turns out to be \cite{TTW1,TTW2},
\begin{equation}
{\rm Prob}(x^+,k^+,k_{\bot}) = \theta(k^+) \theta\left(e A_-(x^+) - k^+\right) 
e^{-2\pi \lambda(k^+,k_{\bot})} \; . \label{prob}
\end{equation}

From the theta functions in (\ref{prob}) we see that particle creation is
an instantaneous and discrete event on the light cone. Why this is so follows
from the fact that evolving in $x^+$ can be regarded as the infinite boost 
limit of evolving in time \cite{Kogut}. Suppose we consider a primed frame 
in which the empty state is specified at $t'=0$. The homogeneous electric 
field will result in particle creation, but the created particles will 
possess finite, nonzero $p^{\prime \pm}$. Now consider the light cone momenta 
in the frame obtained by boosting to velocity $\beta$ in the ${\widehat x}$ 
direction,
\begin{equation}
p^+ = \sqrt{{1 - \beta \over 1 + \beta}} p^{\prime +} \qquad , \qquad
p^- = \sqrt{{1 + \beta \over 1 - \beta}} p^{\prime -} \; .
\end{equation}
As $\beta$ approaches unity we see that $p^+$ goes to zero. Therefore any
particle created in the light cone problem comes out with $p^+ = 0$. But the
physical momentum in a background vector potential is the minimally coupled
one, $p^+ = k^+ - e A_-(x^+)$. Hence pair creation occurs at the instant when
$k^+ = e A_-(x^+)$. Note also that $p^- \rightarrow \infty$ at creation, in
conformity with the eigenvalue we read off from (\ref{eigen}).

We can also compute the instantaneous rate of particle production in $D= 3+1$
dimensions. The probability of still being in vacuum at $x^+$ is \cite{TTW1},
\begin{eqnarray}
\lefteqn{P_{\rm vac}(x^+) = \prod_{0 < k^+ < e A_-} \prod_{k_{\bot}} \left[1 -
e^{-2\pi \lambda(k^+,k_{\bot})}\right]^2 \; ,} \\
& = & \exp\left\{ 2 \int dx^- \int d^2x_{\bot} \int_0^{eA_-} {dk^+ \over 2\pi} 
\int {d^2k_{\bot} \over (2\pi)^2} \ln\left[1 - e^{-2\pi \lambda(k^+,k_{\bot})}
\right]\right\} \; , \\
& = & \exp\left\{ - \frac1{4\pi} \int dx^- \int d^2 x_{\bot} \sum_{n=1}^{
\infty} \left({e E(x^+) \over n \pi}\right)^2 \exp\left[-{n \pi m^2 \over 
\vert e E(x^+) \vert }\right] \right\} \; .
\end{eqnarray}
The particle production rate per 4-volume is minus the logarithmic derivative,
\begin{equation}
- {\partial^4 \ln[P_{\rm vac}(x^+)] \over \partial x^+ \partial x^- \partial^2
x_{\bot}} = {1 \over 4 \pi} \sum_{n =1}^{\infty} \left({e E(x^+) \over n 
\pi}\right)^2 \exp\left[- {n \pi m^2 \over \vert e E(x^+)\vert} \right] \; .
\end{equation}
This has the same {\it form} as Schwinger's result \cite{Schwinger}, but we
have obtained it for an $x^+$ dependent electric field and we have done so
instantaneously, without having to work asymptotically. This is probably the
simplest example we shall ever find of particle production.

\section{The current operators}

It might seem curious that we only see the creation of positrons. In fact one
electron is created for each positron, however, the newly created electrons
immediately leave the light cone manifold. To understand this consider the 
evolution of a virtual pair as depicted in Fig.~2. Electrons are accelerated 
to the speed of light in the negative ${\widehat x}$ direction, which is 
parallel to the $x^-$ axis. They never evolve past a certain value of $x^+$. 
In contrast, positrons are accelerated to the speed of light in the positive 
${\widehat x}$ direction, which is parallel to the $x^+$ axis. We should 
therefore see the $J^+$ operator grow as the manifold fills up with positrons. 
We should also see $J^- \propto (L + x^-)$ from the flux of newly created
electrons originating all the way back to $x^- = -L$.

\centerline{\psfig{figure=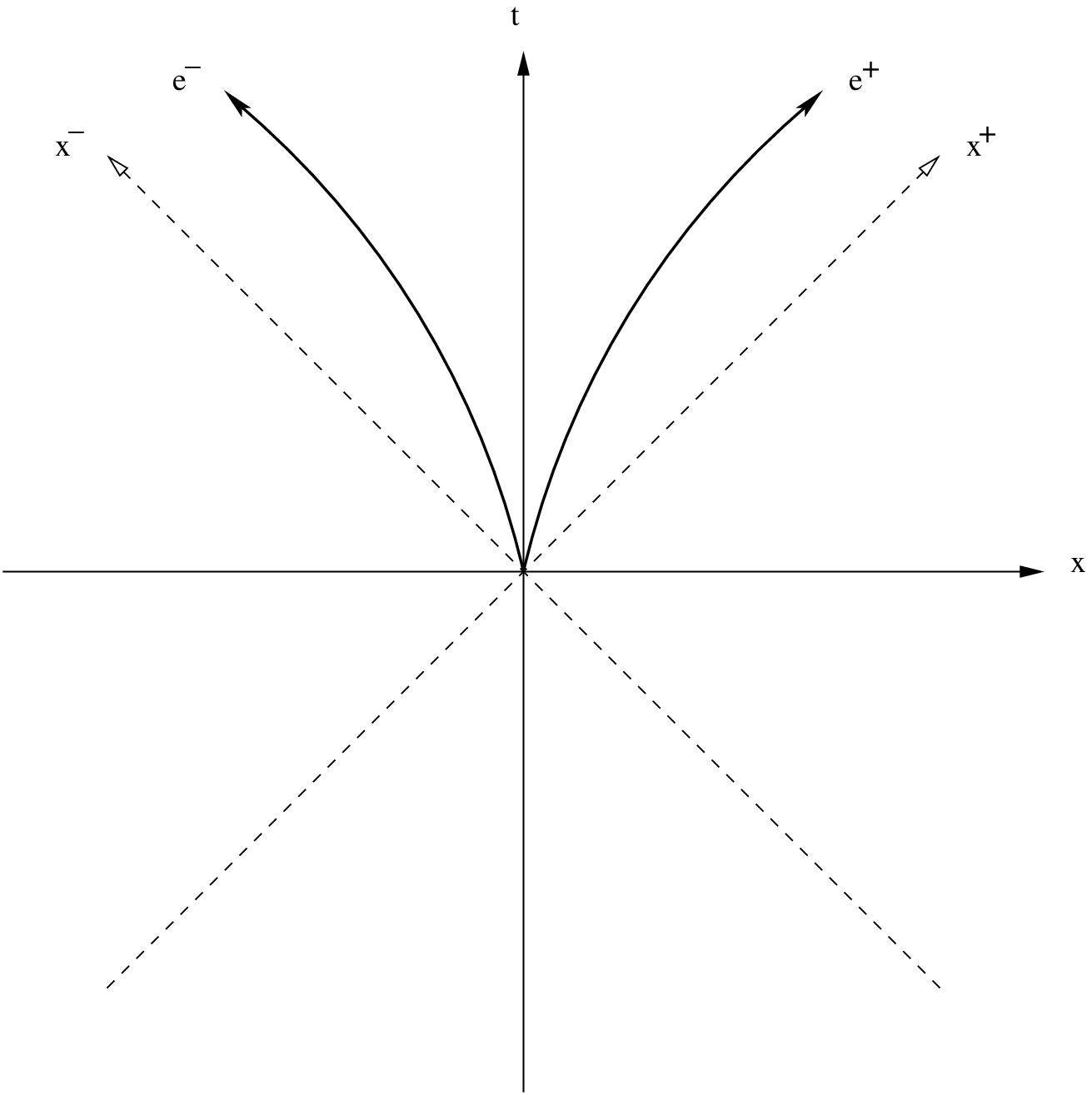,height=3.5in}}

\centerline{Fig.~2: The evolution of a virtual $e^+ e^-$ pair.}
\vskip 1cm

Explicit calculations verify both expectations. The light cone current density 
is $J^+$ and our result for it in $D=3+1$ dimensions is \cite{TTW1},
\begin{equation}
\lim_{L \rightarrow \infty} \Bigl\langle \Omega \Bigl\vert J^+(x^+,x^-,x_{
\bot}) \Bigr\vert \Omega \Bigr\rangle = -2 e \int_0^{eA_-(x^+)} {dp^+ \over
2 \pi} \int {d^2 p_{\bot} \over (2\pi)^2} e^{-2\pi \lambda(p^+,p_{\bot})} 
\; . \label{J+}
\end{equation}
Once one accepts (\ref{prob}) for the creation probability, this result is
simple to understand. Each positron carries charge $-e$, the probability of
creating either spin state with momenta $k^+$ and $k_{\bot}$ is $\exp[-2\pi
\lambda(k^+,k_{\bot})]$, there are two spin states, and the number of modes 
per unit volume is $dp^+/2\pi \times d^2p_{\bot}/(2\pi)^2$. So the differential
increment in the light cone charge density should be,
\begin{equation}
dJ^+ = \Bigl(-e\Bigr) \times \Bigl(e^{-2\pi \lambda(k^+,k_{\bot})}\Bigr)
\times \Bigl(2\Bigr) \times \Biggl({dp^+ \over 2\pi}\Biggr) \times \Biggl(
{d^2p_{\bot} \over (2\pi)^2}\Biggr) \; .
\end{equation}
Our result (\ref{J+}) for $J^+$ is just the integral of this over the modes 
which have undergone pair creation. In exact analogy, we expect the increment
from each element $dx^-$ to the electron flux to be,
\begin{equation}
dJ^- = \Bigl(+e\Bigr) \times \Bigl(e^{-2\pi \lambda(k^+,k_{\bot})}\Bigr)
\times \Bigl(2\Bigr) \times \Biggl({-eE(x^+) dx^- \over 2\pi}\Biggr) \times
\Biggl({d^2p_{\bot} \over (2\pi)^2}\Biggr) \; .
\end{equation}
The total result is the integral from $-L$ to $x^-$, plus the ultraviolet
divergent integration constant. The expectation values of the transverse
currents are zero, so we see that current is indeed conserved.

\section{Axial vector anomaly in $D=1+1$}

To get the current operator VEV's in $D=1+1$ dimensions it is only necessary 
to drop the transverse coordinates, change the number of spin states from two
to one, and drop the ultraviolet divergent integration constant for $J^-$.
The results are \cite{TTW2},
\begin{eqnarray}
\left\langle \Omega \left\vert J^+(x^+,x^-) \right\vert \Omega \right\rangle
= -e \int_0^{eA_-} {dp^+ \over 2\pi} e^{-2\pi \lambda(p^+)} + O\left({\ln(L)
\over L}\right) \; , \\
\left\langle \Omega \left\vert J^-(x^+,x^-) \right\vert \Omega \right\rangle = 
- (L + x^-) {e^2 E(x^+) \over 2\pi} e^{-2\pi \lambda(eA_-)} + O\left({\ln(L)
\over L}\right) \; .
\end{eqnarray}
In $D=1+1$ the components of the axial vector current operator are proportional 
to those of the vector current operator,
\begin{equation}
J_5^{\pm}(x^+,x^-) = \pm {1 \over e} J^{\pm}(x^+,x^-) \; .
\end{equation}
Hence the divergence of the axial vector current is,
\begin{equation}
\lim_{L \rightarrow \infty} \left\langle \Omega \left\vert \partial_+ J^+_5
+ \partial_- J^-_5 \right\vert \Omega \right\rangle = {e^2 E(x^+) \over \pi}
e^{-2 \pi \lambda(eA_-)} \; . \label{div5}
\end{equation}

To complete the computation we need the VEV of the psuedo-scalar, which is
$J_5 = \overline{\psi} \gamma_5 \psi$ up to technicalities of ordering and
regularization. After some straightforward manipulations we obtain \cite{TTW2},
\begin{equation}
\lim_{L \rightarrow \infty} \left\langle \Omega \left\vert J_5(x^+,x^-)
\right\vert \Omega \right\rangle = {i e E(x^+) \over 2 \pi m}
\left[1 - e^{-2 \pi \lambda(eA_-)} \right]\; .
\end{equation}
Combining with (\ref{div5}) we find,
\begin{equation}
\lim_{L \rightarrow \infty} \left\langle \Omega \left\vert \partial_{\mu}
J^{\mu}_5(x^+,x^-) - 2 i e m J_5(x^+,x^-) \right\vert \Omega \right\rangle = 
{e^2 \over \pi} E(x^+) \; .
\end{equation}
As far as I know, this is the first time the axial vector anomaly has been
checked for massive QED in $D=1+1$ dimensions.

To make the same check in $D=3+1$ dimensions one must extend the background 
to include at least a constant magnetic field which is parallel to the 
electric field. The relevant mode functions have been worked out by my graduate
student, Marc Soussa. Evaluation of the various fermion bilinears in the 
presence of this background is already far advanced.

\vskip .5cm
\centerline{\bf Acknowledgments}

It is a pleasure to acknowledge my collaboration on this project with N. C. 
Tsamis and T. N. Tomaras. This work was partially supported by DOE contract 
DE-FG02-97ER\-41029 and by the Institute for Fundamental Theory.


\begin{thebibliography}{99}

\bibitem{Schwinger} J. Schwinger, Phys. Rev. 82, 664 (1951).

\bibitem{TTW1} N. T. Tomaras, N. C. Tsamis and R. P. Woodard, Phys. Rev. {\bf
D62}, 125005, 2000, hep-ph/0007166.

\bibitem{TTW2} N. T. Tomaras, N. C. Tsamis and R. P. Woodard, ``Pair creation
and axial vector anomaly in light-cone $QED_2$,'' to appear in JHEP,
hep-th/0108090.

\bibitem{Kogut} J. B. Kogut and D. E. Soper, Phys. Rev. {\bf D1}, 2901 (1970).

\end{thebibliography}
\end{document}